\documentclass[a4paper,twoside]{article}

\usepackage[utf8]{inputenc}
\usepackage{ifthen}
\usepackage{amssymb}
\usepackage{amsmath}
\usepackage{amstext}
\usepackage{amsthm}
\usepackage{booktabs}
\usepackage{multirow}
\usepackage{graphicx}
\usepackage{subfigure}
\usepackage{pslatex}
\usepackage{apalike}
\usepackage{SCITEPRESS}  

\graphicspath{{figures/}}

\newboolean{blind}
\setboolean{blind}{false}

\begin{document}

\title{Classification of Cardiac Arrhythmias from Single Lead ECG \\
    with a Convolutional Recurrent Neural Network}

\ifthenelse{\boolean{blind}}{}{
\author{
\authorname{
    Jérôme~Van~Zaen,
    Olivier~Chételat,
    Mathieu~Lemay, \\
    Enric~M.~Calvo, and
    Ricard~Delgado-Gonzalo}
\affiliation{
    Swiss Center for Electronics and Microtechnology (CSEM),
    Rue Jaquet-Droz 1, Neuchâtel, Switzerland}
\email{\{jerome.vanzaen, olivier.chetelat, mathieu.lemay, enric.muntanecalvo,
    ricard.delgado\}@csem.ch}
}
}


\abstract{
While most heart arrhythmias are not immediately harmful, they can lead to
severe complications. In particular, atrial fibrillation, the most common
arrhythmia, is characterized by fast and irregular heart beats and increases
the risk of suffering a stroke. To detect such abnormal heart conditions, we
propose a system composed of two main parts: a smart vest with two cooperative
sensors to collect ECG data and a neural network architecture to classify heart
rhythms. The smart vest uses two dry bi-electrodes to record a single lead ECG
signal. The biopotential signal is then streamed via a gateway to the cloud
where a neural network detects and classifies the heart arrhythmias. We
selected an architecture that combines convolutional and recurrent layers. The
convolutional layers extract relevant features from sliding windows of ECG and
the recurrent layer aggregates them for a final softmax layer that performs the
classification. Our neural network achieves an accuracy of 87.50\% on the
dataset of the challenge of Computing in Cardiology 2017.
} 

\onecolumn \maketitle \normalsize \vfill

\section{\uppercase{Introduction}}
\label{sec:introduction}

\noindent
Heart arrhythmias are caused by irregular electrical conduction in cardiac
tissue. Atrial fibrillation, which affects 1--2\% of the population~\cite{Camm2010},
is the most common one. Furthermore, its prevalence increases
with age, from $<$0.5\% at 40--50 years to 5--15\% at 80 years. While not
directly life-threatening, it can lead to serious complications~\cite{January2014}.
In particular, atrial fibrillation is associated with a
3--5 fold increased risk of stroke and a 2-fold increased risk of mortality~\cite{Kannel1998}.
It is also associated with a 3-fold risk of heart failure~\cite{Wang2003}.
Typical symptoms include heart palpitations, shortness of
breath, and fainting. However, about one third of the cases are asymptomatic
which prevents early diagnosis. This, in turn, precludes early therapies which
might protect the patient from the consequences of atrial fibrillation but also
from its progression. Indeed, atrial fibrillation causes electrical and
structural remodeling of the atria which facilitates its further
development~\cite{Frick2001,Nattel2008}.

The gold standard for diagnosing atrial fibrillation and other heart
arrhythmias is the 12-lead ECG. A trained electrophysiologist can select the
most appropriate treatment after reviewing ECG signals and the patient history.
This is, however, a time-consuming task, especially for long recordings such as
the ones collected with Holter monitors. To alleviate this task, several
approaches have been proposed to detect arrhythmias from ECG
signals~\cite{Owis2002,DeChazal2004}. Even without perfect detection accuracy, these
approaches are useful as they facilitate reviewing ECG by selecting relevant
signal excerpts.

Recently, neural networks have shown impressive performance in various
classification and regression tasks. Image processing was the first field where
deep networks surpassed existing approaches by a large margin~\cite{Krizhevsky2012}.
Since then, they have been extensively applied to fields
previously dominated by signal processing. In particular, several architectures
have been proposed to detect and classify heart arrhythmias from ECG signals.

In the context of the challenge of Computing in Cardiology 2017~\cite{Clifford2017},
a few neural network architectures were proposed to
classify single lead ECG signals into one of the following classes: normal
sinus rhythm, atrial fibrillation, other rhythm, and noise. One of these
architectures uses logarithmic spectrograms computed over sliding windows of
ECG as input to two-dimensional convolutional layers~\cite{Zihlmann2017}.
Aggregation of successive windows is done with either temporal averaging or a
recurrent layer. However, the convolutional and recurrent layers were trained
separately due to convergence issues. A similar approach used a 16-layer
convolutional neural network with skip connections to classify arrhythmias from
ECG signals~\cite{Xiong2017}. Each layer is composed of batch normalization,
ReLU activation, dropout, one-dimensional convolution, and averaging pooling.

Recently, a convolutional neural network was shown to reach cardiologist-level
arrhythmia detection~\cite{Rajpurkar2017}. This 34-layer network takes
advantage of a very large dataset of 64,121 ECG signals, recorded from 29,163
patients, to recognize 12 different heart arrhythmias including atrial
fibrillation, atrial flutter, and ventricular tachycardia. Another approach
applied convolutional neural networks for detecting atrial fibrillation from
time-frequency representations of ECG signals~\cite{Xia2018}. Two methods for
computing these representations were compared: the short-time Fourier transform
or the stationary wavelet transform. In this case, the neural network using
coefficients from the second transform yielded better classification accuracy.

Neural networks have thus shown promising results for the detection of abnormal
cardiac rhythms. Furthermore, as mentioned previously, it is of the utmost
importance to detect arrhythmias as early as possible to improve treatment
outcome. To tackle this issue, we developed a system composed of two main
elements: a smart vest to record ECG signals and an algorithm to detect and
classify arrhythmias. The smart vest includes two cooperative sensors to record
a single lead ECG signal and stream by Bluetooth the collected data to a
gateway. This gateway then forwards the ECG signal to the cloud where it is
processed by a neural network in order to detect abnormal rhythms.

This article is structured as follows. First, the dataset of ECG signals, our
neural network architecture, and the monitoring system are described in
Section~\ref{sec:methods}. Then, the results are presented in
Section~\ref{sec:results} and discussed in Section~\ref{sec:discussion}.
Finally, this article ends with a short conclusion in
Section~\ref{sec:conclusion}.

\section{\uppercase{Materials and Methods}}
\label{sec:methods}

\noindent

\subsection{Dataset}

We used the dataset from the challenge of Computing in Cardiology 2017 to train
a neural network to classify cardiac arrhythmias. This dataset includes 8528
single lead ECG signals recorded with an AliveCor device. The signals are
sampled at 300~Hz and have durations ranging from 9 to 60 seconds. Each signal
was acquired when the subject held each one of the two electrodes in each hand
resulting in a lead I (left arm -- right arm) ECG. As the device has no
specific orientation, many signals are inverted (right arm -- left arm).

All ECG signals are labeled with one of the following four classes: normal
sinus rhythm, atrial fibrillation, other rhythm, and noise. The proportion of
each class in the dataset varies from 3.27\% for noise to 59.52\% for normal
rhythm. The full breakdown of all classes is reported in
Table~\ref{table:breakdown}.

\begin{table}
    \centering
    \scriptsize
    \caption{Number of signals and mean duration for each class.}
    \renewcommand{\arraystretch}{1.3}
    \begin{tabular}{@{}lrrr@{}}
    \toprule
    Class & Count & Proportion & Mean duration [s] \\
    \midrule
    Normal rhythm & 5076 & 59.52\% & 32.11 \\
    Atrial fibrillation & 758 & 8.89\% & 32.34 \\
    Other rhythm & 2415 & 28.32\% & 34.30 \\
    Noise & 279 & 3.27\% & 24.38 \\
    Total & 8528 & 100\% & 32.50 \\
    \bottomrule
    \end{tabular}
    \label{table:breakdown}
\end{table}

The entries of the challenge of Computing in Cardiology 2017 were ranked
according to the following score evaluated on a private test set:
\begin{equation}
    S_{\text{CinC}} = \frac{F_{1n} + F_{1a} + F_{1o}}{3} \label{eq:cinc2017}
\end{equation}
where $F_{1n}$, $F_{1a}$, and $F_{1o}$ denote the $F_1$ scores for normal
rhythm, atrial fibrillation, and other rhythm. The four
winners~\cite{Teijeiro2017,Datta2017,Zabihi2017,Hong2017} reached a score of 0.83.

\begin{figure}
    \centering
    \includegraphics[width=75mm]{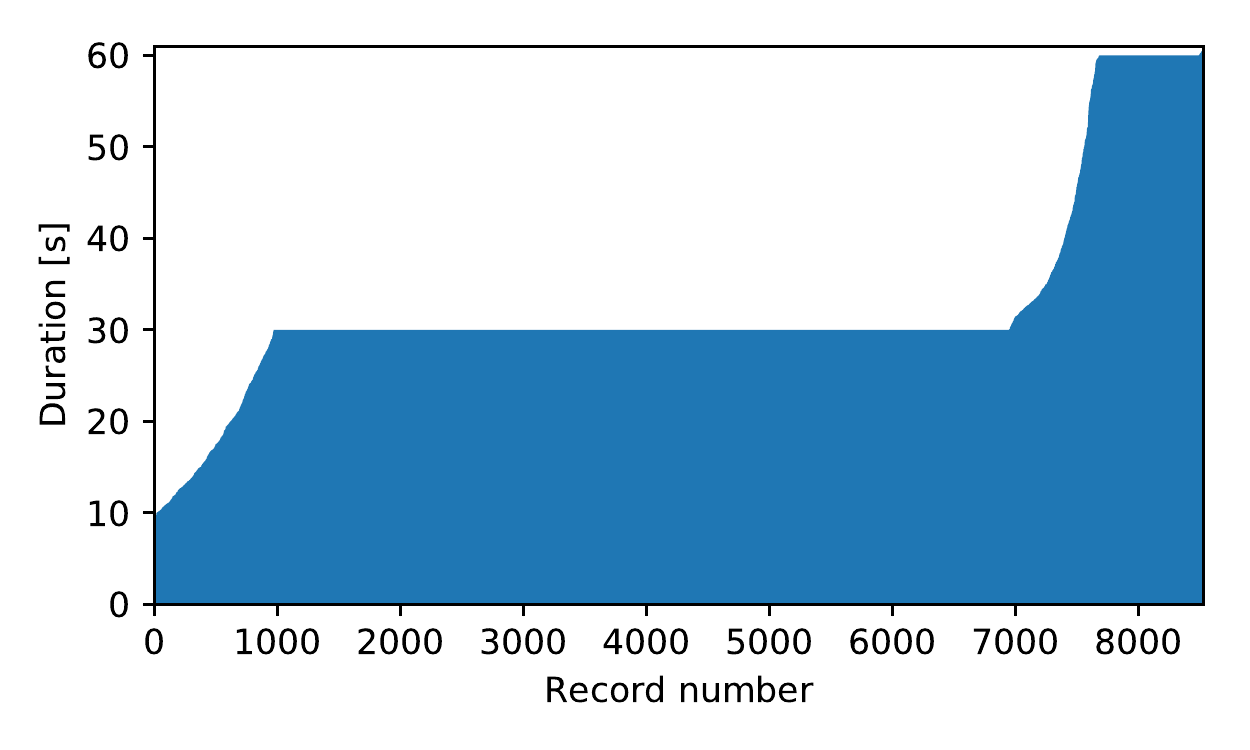}
    \caption{Record durations sorted in increasing order.}
    \label{fig:durations}
\end{figure}

\begin{figure}
    \centering
    \includegraphics[width=75mm]{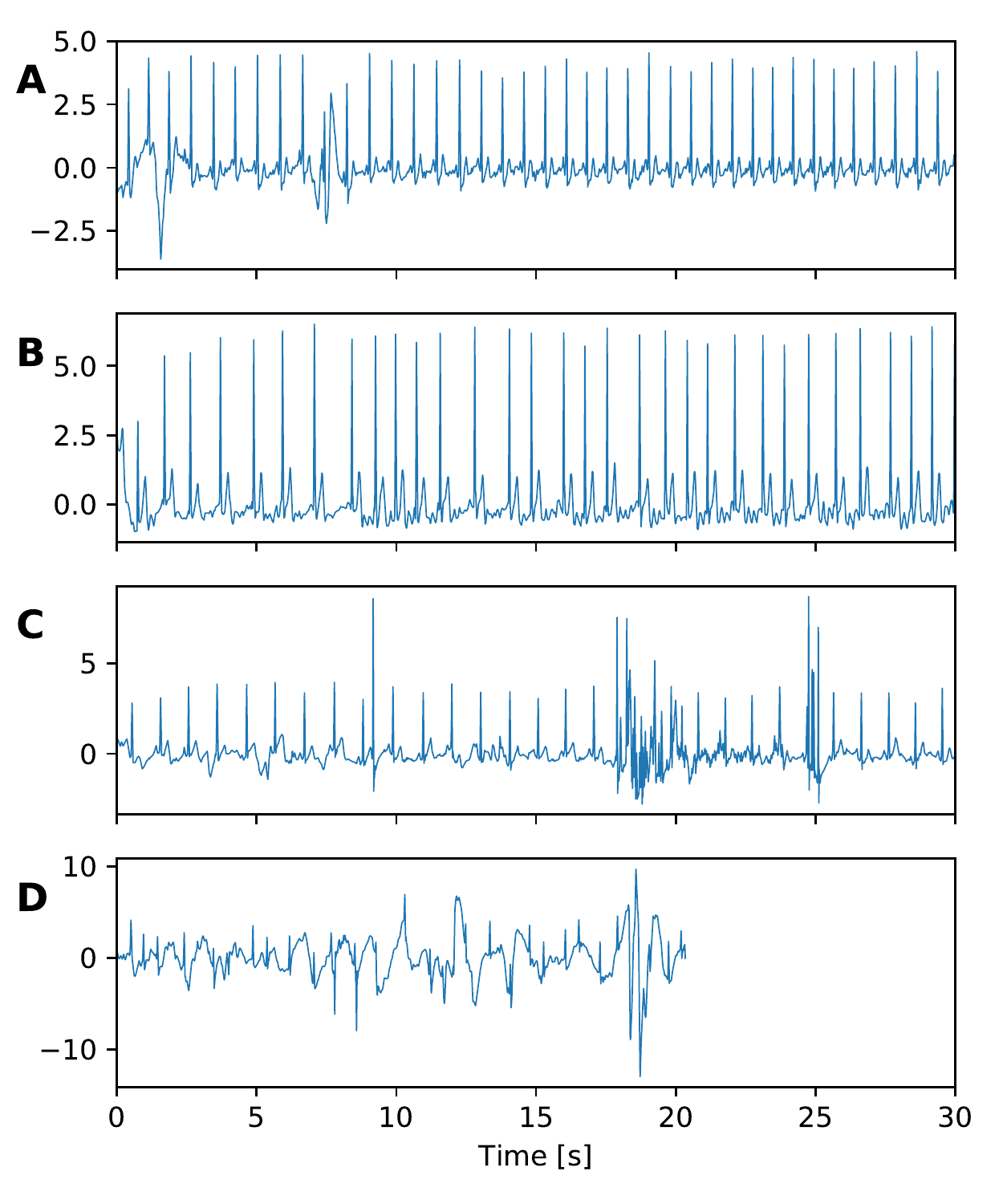}
    \caption{Examples of ECG records: (A) normal rhythm from record A00001, (B)
        atrial fibrillation from record A00004, (C) normal rhythm from record
        A00002, (D) atrial fibrillation from record A00015.}
    \label{fig:examples}
\end{figure}

Several aspects of this dataset are challenging for arrhythmia classification.
First, many signals are inverted as mentioned previously. Second, the classes
are not balanced. There are very few signals labeled atrial fibrillation or
noise compared to the ones labeled normal rhythm. Furthermore, the durations of
the recordings are also different. They vary from 9 to 60 seconds.
Figure~\ref{fig:durations} illustrates this issue. Most ECG signals last 30
seconds but a significant portion has shorter or longer durations. In addition,
labeling is relatively coarse as a single label is associated with each ECG
signal. In some cases, several labels could be used for the same signal.
Finally, the ECG quality of a non-negligible part of the records is rather
poor. Four examples of signals are shown in Figure~\ref{fig:examples}. The
first two signals are labeled normal rhythm and atrial fibrillation and have
good overall quality. The third example is a normal rhythm record with
acceptable quality except for a short segment of noise. The last ECG signal is
labeled as atrial fibrillation but has very poor quality due to large shifts of
the baseline. It also illustrates that all signals do not share the same
duration.

The test set used during the challenge of Computing in Cardiology was not yet
publicly released. Therefore, we split the dataset (which was originally
intended only for training) into a training set of 7000 records and a test set
with the remaining 1528 records while approximately preserving class
repartition.

We applied the following pre-processing steps to the dataset. First, we
filtered the ECG signals with a digital Butterworth band-pass filter between
0.5 and 40~Hz. The filter is applied in both forward and backward directions to
avoid distortions. The analog filter used by the recording device has similar
cutoff frequencies but still leaked some components outside the pass-band.
Then, we downsampled the signals to 200~Hz to reduce the number of samples.
Finally, the signals are scaled by the mean of the standard deviations of all
signals from the training set. Scaling is helpful to accelerate training~\cite{LeCun2012}.

\subsection{Network Architecture}

An approach to handle signals with different durations is to extract windows
with the same length. The label of a signal is then used for all included
windows. Fixed size inputs would make possible to apply a convolutional neural
network to learn features useful to classify arrhythmias. However, this
approach is sub-optimal as illustrated in Figure~\ref{fig:examples} where a
signal labeled as normal rhythm includes a segment of noise. Furthermore, all
signals would need to be truncated to the same length. This would lead to a
large data loss due to the considerable variations in signal durations.

A more appropriate approach is to use a recurrent neural network which is
well-suited to process sequences with varying lengths. While such a neural
network can, by design, remember past values over long time intervals, they are
not as efficient as convolutional neural networks for learning complex
features.

After reviewing the advantages and drawbacks of these two approaches, we chose
to combine them and build a neural network that includes convolutional and
recurrent layers. Specifically, each ECG signal was divided in sliding windows
of 512 samples with 50\% overlap. This corresponds to a window duration
slightly above 2.5 seconds. The number of windows extracted from each signal
depends on its duration. Seven convolutional layers were applied to all windows
of a signal. Each convolutional layer is composed of a 1D convolution and a
max pooling operation. The convolution uses a kernel of size 5, zero padding,
and a ReLU activation~\cite{Hahnloser2000}. The pool size for max pooling was
set to 2. The first convolutional layer has 8 output channels (from the single
channel windows). Then, each following layers double the number of channels
while max pooling halves the number of samples. After the convolutional layers,
a global averaging pooling layer was applied. This results in 512 features for
each input window. The features were then processed with a long short-term
memory (LSTM) layer~\cite{Hochreiter1997} with 128 units. Finally, a softmax
layer outputs the probability of each class for the input ECG windows.
This results in a neural network with 1,203,364 trainable parameters including
874,656 for the convolutional part, 328,192 for the recurrent part, and 516 for
the final softmax layer. The complete network architecture is summarized in
Table~\ref{table:architecture}.

\begin{table}
    \centering
    \caption{Neural network architecture. The output size is given as
        $N_w \times N_s \times N_c$ where $N_w$ denotes the variable number of
        windows, $N_s$ the number of samples, and $N_c$ the number of
        channels.}
    \renewcommand{\arraystretch}{1.3}
    \begin{tabular}{@{}lc@{}}
        \toprule
        Layer & Output size \\
        \midrule
        Input windows & $N_w \times 512 \times 1$ \\
        Convolutional layer 1 & $N_w \times 256 \times 8$ \\
        Convolutional layer 2 & $N_w \times 128 \times 16$ \\
        Convolutional layer 3 & $N_w \times 64 \times 32$ \\
        Convolutional layer 4 & $N_w \times 32 \times 64$ \\
        Convolutional layer 5 & $N_w \times 16 \times 128$ \\
        Convolutional layer 6 & $N_w \times 8 \times 256$ \\
        Convolutional layer 7 & $N_w \times 4 \times 512$ \\
        Global average pooling & $N_w \times 512$ \\
        LSTM layer & $128$ \\
        Softmax layer & $4$ \\
        \bottomrule
    \end{tabular}
    \label{table:architecture}
\end{table}

\subsection{Data Augmentation}

As the dataset is relatively small for fitting a neural network, we applied
different strategies to synthetically augment the number of ECG signals
available during training. The first strategy is to simply flip the sign of
each signal with probability 0.5. We found it easier to let the neural network
learn to take into account inverted ECG signals instead of applying a
rectifying step during pre-processing.

Furthermore, when extracting sliding windows, it is not possible to use all
samples for the large majority of ECG signals. Indeed, the maximum number of
sliding windows $N_w$ in a signal with $N$ samples is given by
\begin{displaymath}
    N_w = \left\lfloor \frac{N - 512}{256} \right\rfloor + 1
\end{displaymath}
assuming $N \ge 512$. In the previous expression, $\lfloor\cdot\rfloor$ denotes
the floor function. We took advantage of this fact to place the first window
at a random offset from the start of the signal. This random offset is drawn
uniformly from
\begin{displaymath}
    \{0, 1, \dots, N - (N_w - 1) \cdot 256 - 512\}
\end{displaymath}
for each signal at each epoch. The main idea behind this strategy is to prevent
the neural network from learning the exact positions of the QRS complexes in
the training set. However, we always used the maximum number of sliding windows
possible for each signal to avoid wasting ECG samples.

The third strategy we applied is to resample each signal at each epoch with
probability 0.8 in order to simulate slightly slower or faster heart rate and
thus help the neural network to reach better generalization performance.
Naturally, the resampling operation should not change the heart rate too much.
Otherwise, there is a risk to confuse a cardiac rhythm for another one.
Therefore, if a signal needs resampling, its length is changed by a proportion
sampled uniformly between $-5$\% and $+5$\%.

\subsection{Training}

We implemented our neural network and strategies for data augmentation in
Python with the Keras library~\cite{Chollet2015}. We trained the neural network
for 200 epochs by minimizing the cross-entropy with the Adam
algorithm~\cite{Kingma2014}. We set the initial learning rate to 0.001. The learning rate
was divided by two if the cross-entropy evaluated on the test set did not
decrease for 5 consecutive epochs with a lower limit at $10^{-5}$.

We used a batch size of 50 signals. As a batch must include the same number of
sliding windows for each signal, we applied zero-padding. Specifically, too
short signals were prepended with all-zero windows. To limit zero-padding as
much as possible we sorted the signals by duration and grouped them in batches
of similar lengths. This resulted in batches with varying numbers of windows.

The LSTM layer was regularized  by applying dropout with a rate of 0.5 for both
the input and recurrent parts~\cite{Srivastava2014,Gal2016}. We monitored the
accuracy on the test set and selected the weights at the best epoch as the
final parameters of the neural network.

\subsection{Monitoring System}

\begin{figure}
    \centering
    \ifthenelse{\boolean{blind}}{
        \includegraphics[width=75mm]{pixelized_sense_sensors.jpg}
    }{
        \includegraphics[width=75mm]{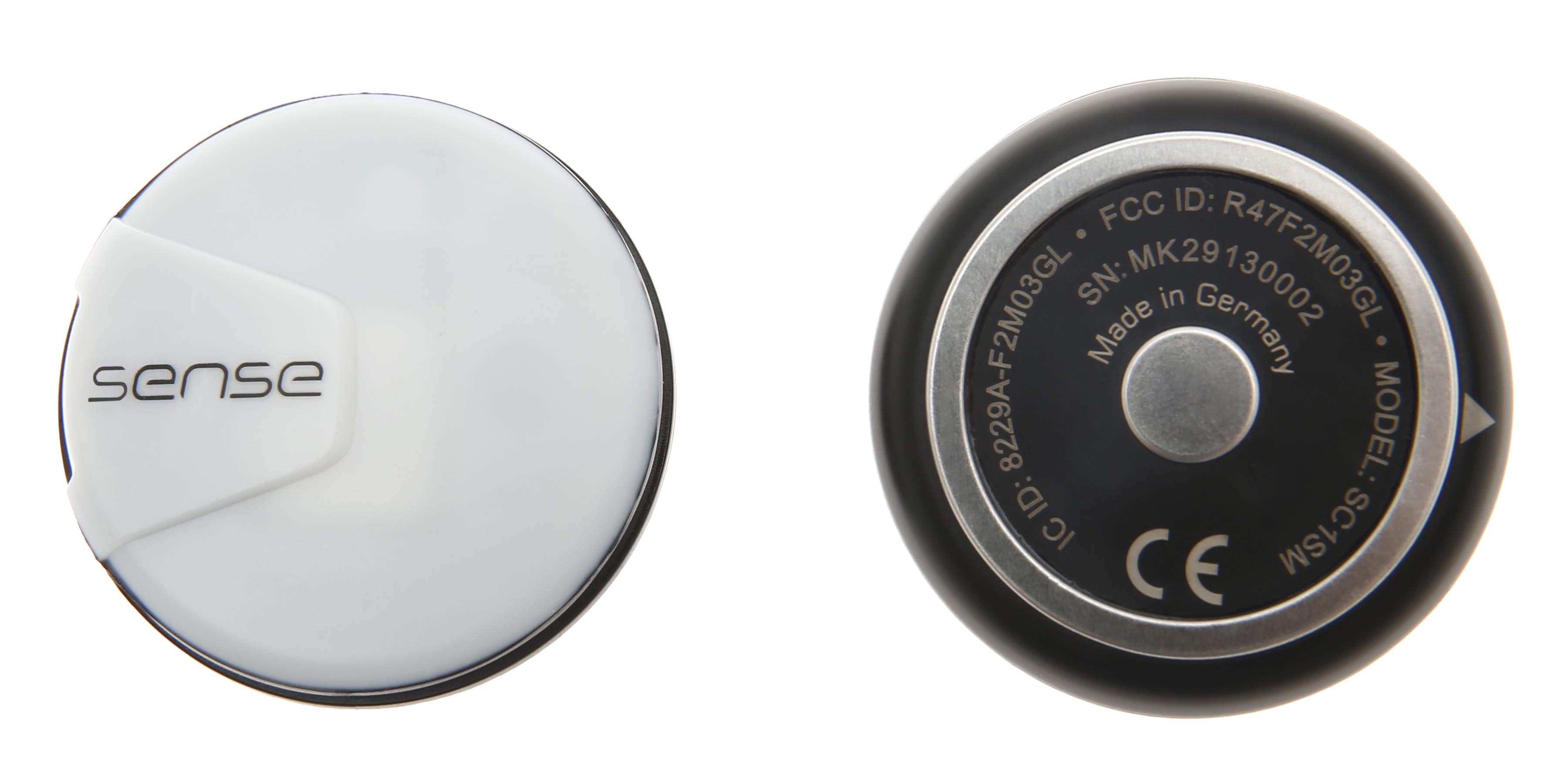}
    }
    \caption{Cooperative sensors included in the smart vest for monitoring
        ECG.}
    \label{fig:sense_sensors}
\end{figure}

The smart vest used to monitor ECG includes two cooperative sensors
illustrated in Figure~\ref{fig:sense_sensors}. These sensors use dry stainless
steel bi-electrodes. This technology, which was validated in a previous study,
yields high quality measurements of ECG and bio-impedance signals, even in
motion. No wetting of the electrode is required. Moreover, the electrical
connection linking both sensors does not have to be shielded, nor insulated,
which makes its integration in garment easier and cheaper (a conductive fabric
is sufficient). The length of the connection is not limited to very short
distances and is therefore placed in the back so as to have space for a central
zipper in the vest, which makes donning and doffing easy, even for the elderly.
The two watertight sensors are clipped in the vest and can be removed for
washing and recharging. Operation is simplified to its maximum: the sensors
automatically switch to record mode as soon as they are applied on the skin and
return to standby mode when removed.

\begin{figure}
    \centering
    \ifthenelse{\boolean{blind}}{
        \includegraphics[width=75mm]{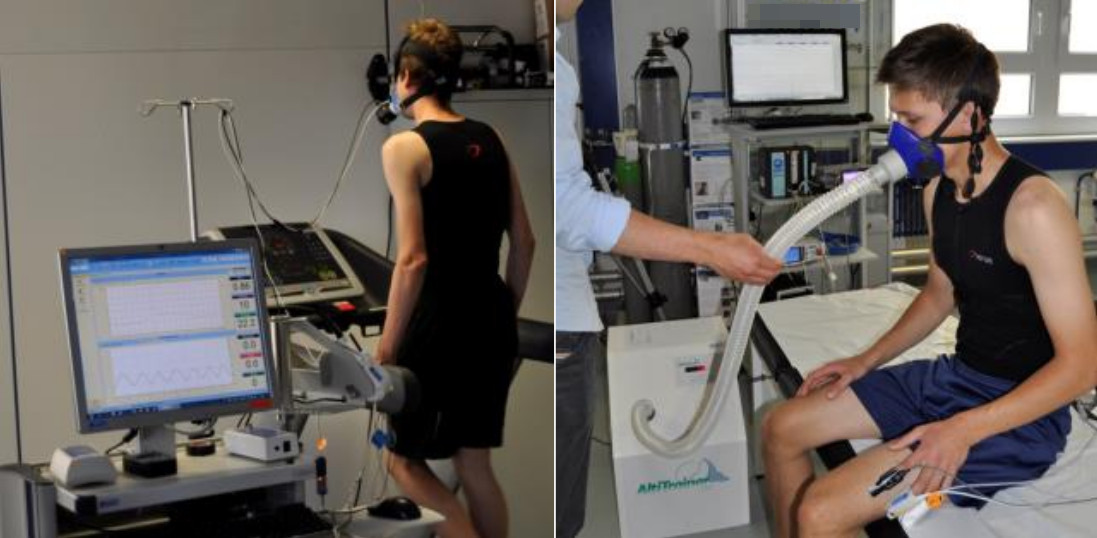}
    }{
        \includegraphics[width=75mm]{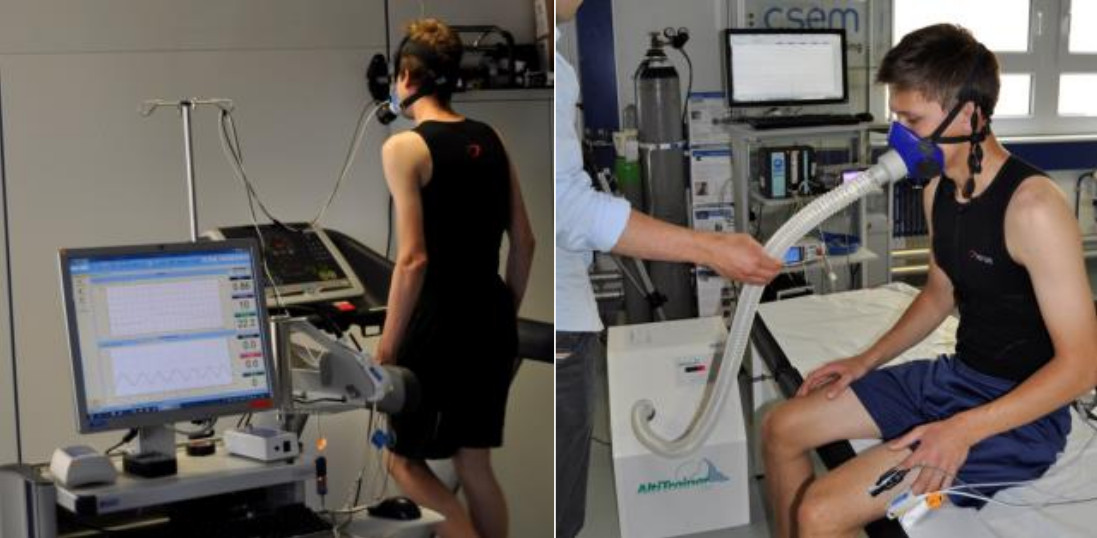}
    }
    \caption{Smart vest during the validation protocol.}
    \label{fig:sense_vest}
\end{figure}

This smart vest which was originally developed for athletes also monitors the
following biomedical signals: heart rate, transthoracic impedance, respiration
rate, skin temperature, activity class (resting, walking, running), and posture
(lying, standing/sitting). A quality index is associated with each signal so
that the reliability of the measurements can be easily assessed. However, these
additional signals and the quality indices were not used in the present study.
Figure~\ref{fig:sense_vest} shows the smart vest during the validation
protocol.

Another important feature of the smart vest consists in its capability to store
all the recorded signals locally as well as synchronize them in the cloud. This
is achieved by means of an accompanying gateway that acts as a relay for the
streamed data. The gateway is robust to poor network connectivity and uploads
the newly available data to the cloud as it is being streamed while an active
connection is available. In our case, the gateway is composed of a simple
Raspberry Pi that connects to the smart vest via Bluetooth. The biomedical data
is then collected, stored locally in a compact format and relayed via telemetry
messages simultaneously to all the clients connected to this gateway as well as
a private cloud. The messaging protocol chosen for this application is
MQTT\footnote{https://mqtt.org} (Message Queuing Telemetry Transport). This
protocol, which uses the publish/subscribe paradigm, is one of the most widely
employed telemetry protocols in IoT and real-time streaming. Once the ECG
signal is in the cloud, we can leverage the powerful computing capabilities and
detect heart arrhythmias with our neural network.

The ECG signals recorded with the smart vest device were pre-processed
similarly to the data from the challenge of Computing in Cardiology 2017.
First, the same band-pass filter between 0.5 and 40~Hz was used to remove the
baseline and high-frequency noise. Then, the signals were resampled at 200~Hz.
Finally, we scaled each signal with its standard deviation. Indeed, we could
not use the scaling factor computed on the training set as the two types of
ECG signals did not have the same range of values.

After pre-processing the ECG signals, we extracted sliding windows of 512
samples with 50\% overlap. We used groups of 25 such windows as input to the
neural network. We selected this specific number of windows as it corresponds
to segments of approximately 33 seconds which is close to the median length of
the signals used for training and testing the model. Applying the neural
network resulted in a rhythm prediction for each group of 25 windows.

\section{\uppercase{Results}}
\label{sec:results}

\begin{figure}
    \centering
    \includegraphics[width=75mm]{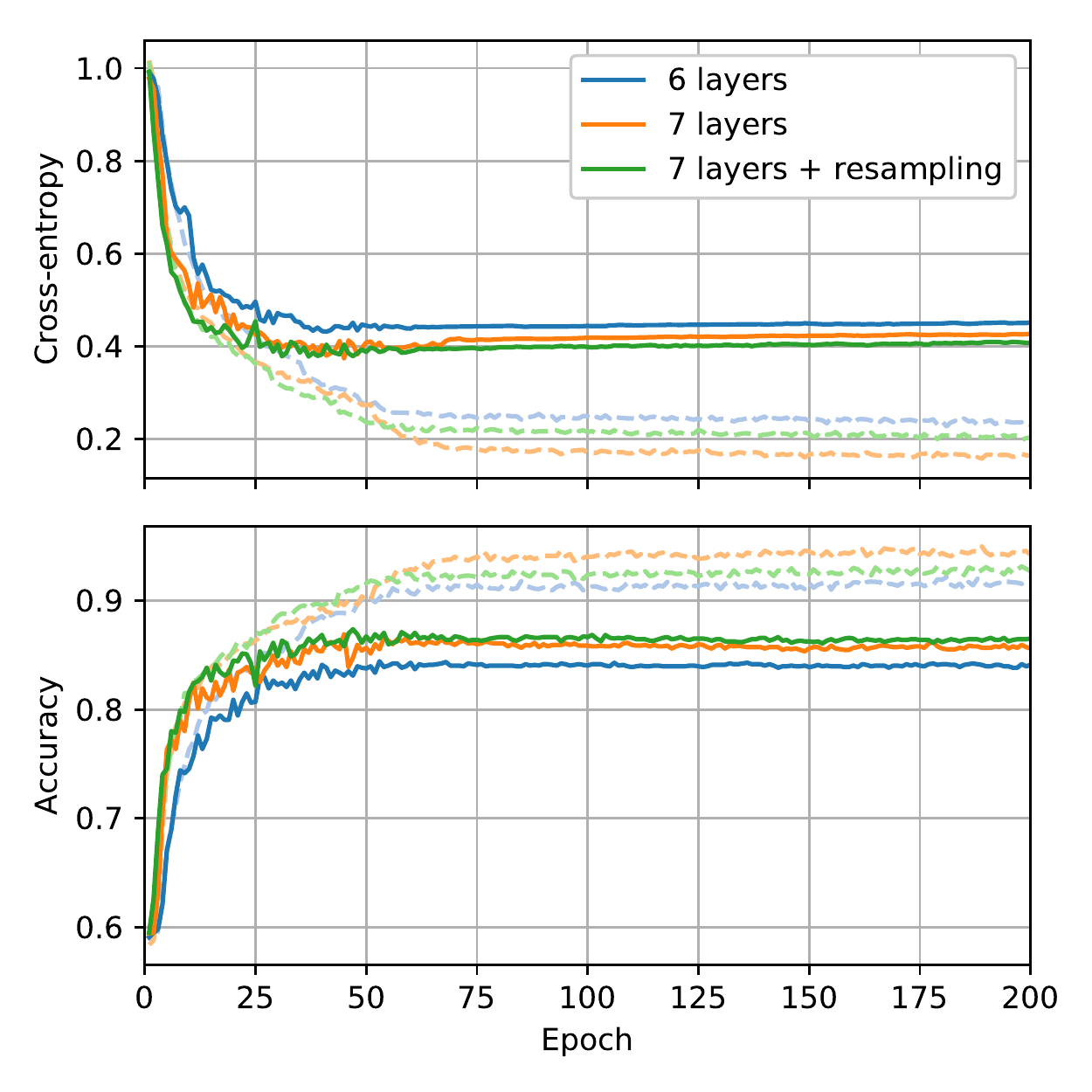}
    \caption{Cross-entropy loss (top) and accuracy (bottom) evaluated during
        training for a network with 6 convolutional layers without resampling
        (blue), a network with 7 convolutional layers without resampling
        (orange), and a network with 7 convolutional layers with resampling
        (green). Dashed lines denote results obtained on the training set and
        solid lines results obtained on the test set.}
    \label{fig:training}
\end{figure}

\noindent
We evaluated three configurations. The first one used the architecture
described in Table~\ref{table:architecture} except that it included only 6
convolutional layers. In addition, all data augmentation strategies were
applied during training except resampling. The second configuration used the
full architecture with 7 convolutional layers but again without resampling.
Finally, the last configuration was identical to the second one but with
resampling for further data augmentation. Figure~\ref{fig:training} shows the
evolution of the cross-entropy loss and the accuracy during training for these
three network configurations. An additional convolutional layer helped to
increase the accuracy and reduce the loss. Resampling the signals also slightly
improved performance. Despite our efforts, we could not completely eliminate
over-fitting as shown by the performance gap between training and test sets.
Indeed, additional regularization only decreased the performance of the
network. However, it is worth mentioning that resampling helped to reduce
over-fitting. The neural network with the third configuration reached an
accuracy of 87.37\% on the test set after 47 epochs. At the same epoch, the
accuracy on the training set was 90.90\%.

After selecting the best neural network, we evaluated it without zero-padding
by selecting a batch size of 1. In this case, the accuracy was 92.64\% on the
training set and 87.50\% on the test set. Sensitivity, specificity, and
$F_1$ score values are reported in Table~\ref{table:metrics} for all classes.
Unsurprisingly, the best $F_1$ score was obtained for the class with the most
samples (normal rhythm) and the worst one for the class with the least samples
(noise). Interestingly, the specificity for atrial fibrillation was relatively
high at 0.9784 while the sensitivity was lower at 0.8382. Thus, the number of
false positive is more larger than the number of false negative which is an
important property of the model. Indeed, it missed only a few atrial
fibrillation cases while false detections can always be disproved with
additional analyses such as a full 12-lead ECG. Furthermore, when evaluated in
terms of the score used during the challenge of Computing in Cardiology 2017
\eqref{eq:cinc2017}, the neural network yielded 0.9156 and 0.8495 on the
training and test sets. This is similar to the best score obtained by the
winners of the challenge. However, we could only evaluate our neural network on
1528 ECG signals from the original training set since the official test set was
not made public yet. Therefore, it is difficult to compare the performance of
our approach with the winning entries.

\begin{table}
    \centering
    \small
    \caption{Performance metrics for the best neural network.}
    \renewcommand{\arraystretch}{1.3}
    \begin{tabular}{@{}llcc@{}}
        \toprule
        Class & Metric & Training set & Test set \\
        \midrule
        \multirow{3}{*}{Normal rhythm} & Sensitivity & 0.9707 & 0.9263 \\
        & Specificity & 0.8955 & 0.8853 \\
        & $F_1$ score & 0.9509 & 0.9243 \\
        \addlinespace
        \multirow{3}{*}{Atrial fibrillation} & Sensitivity & 0.9084 & 0.8382 \\
        & Specificity & 0.9942 & 0.9784 \\
        & $F_1$ score & 0.9232 & 0.8143 \\
        \addlinespace
        \multirow{3}{*}{Other rhythm} & Sensitivity & 0.8380 & 0.7921 \\
        & Specificity & 0.9675 & 0.9352 \\
        & $F_1$ score & 0.8728 & 0.8099\\
        \addlinespace
        \multirow{3}{*}{Noise} & Sensitivity & 0.9345 & 0.7600 \\
        & Specificity & 0.9972 & 0.9871 \\
        & $F_1$ score & 0.9264 & 0.7103 \\
        \bottomrule
    \end{tabular}
    \label{table:metrics}
\end{table}

\begin{figure*}
    \centering
    \includegraphics[width=155mm]{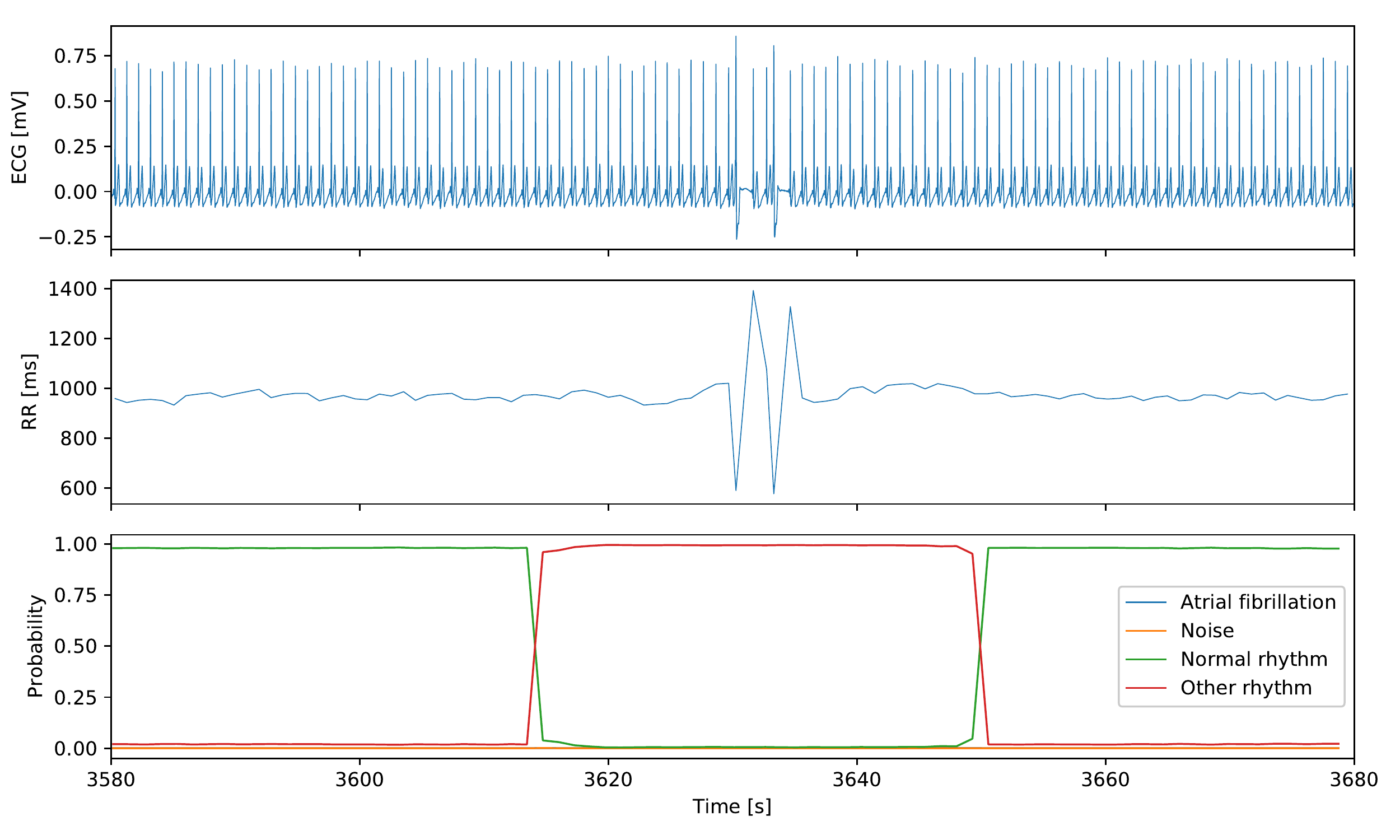}
    \caption{Cardiac rhythm classification from an ECG signal recorded with the
        smart vest. The ECG signal (top) includes two premature ventricular
        contractions which are visible in the RR intervals (middle). The neural
        network correctly identifies this segment of the signal as reflected by
        the class probabilities (bottom).}
    \label{fig:sense_normal_other_normal}
\end{figure*}

After training and evaluation on the dataset of the challenge of Computing in
Cardiology 2017, we applied the neural network to a few ECG signals recorded
with the smart vest. We also extracted the times between consecutive R-waves
from the ECG signals (a.k.a. RR intervals) to facilitate visualization of the
heart rate. An example is shown in Figure~\ref{fig:sense_normal_other_normal}.
In this case, the subject had a normal sinus rhythm with two successive
premature ventricular contractions characterized a very short RR interval
followed by a long recovery RR interval. These premature contractions are
clearly identifiable especially compared to the stable RR intervals of a normal
rhythm. The neural network classified the this ECG segment as normal rhythm
except for the groups of windows including the premature contractions which are
classified as other rhythm.

\begin{figure*}
    \centering
    \includegraphics[width=155mm]{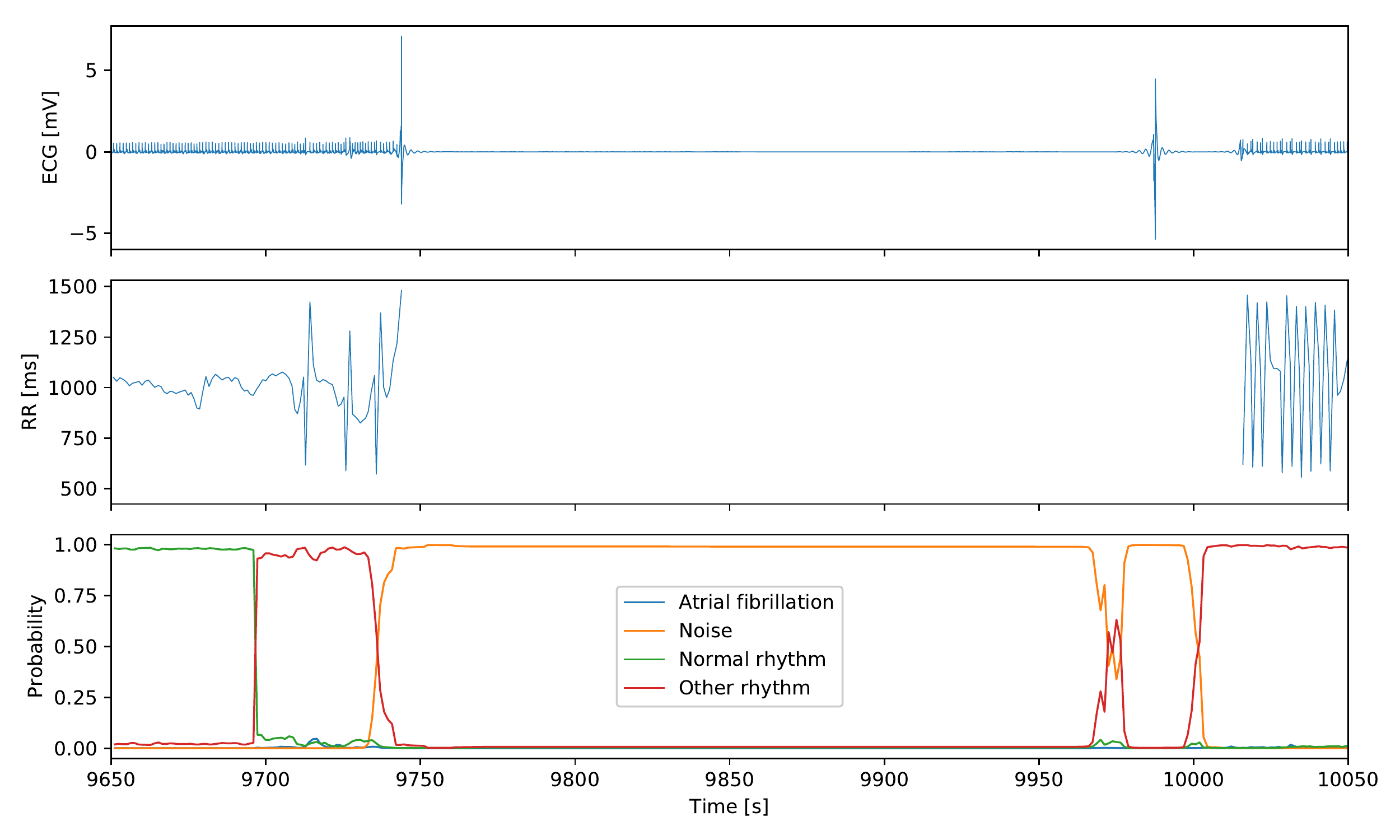}
    \caption{Cardiac rhythm classification from an ECG signal recorded with the
        smart vest. The ECG signal (top) includes a segment with no R waves due
        to poor skin-electrode contact. Consequently, the RR intervals (middle)
        could not be extracted. The class probabilities (bottom) computed by
        the neural network are valid as the segment is labeled as noise.}
    \label{fig:sense_noise}
\end{figure*}

Another example of classification is shown in Figure~\ref{fig:sense_noise}. In
this case, no ECG signal was collected during a few minutes due to poor
skin-electrode contact caused by motion. The neural network identified this
segment without signal as noise. Furthermore, two segments preceding signal
loss were classified as normal rhythm and other rhythm. The RR intervals were
stable in the first segment and there were a few premature ventricular
contractions in the second one. After the ECG signal was recovered, the RR
intervals varied widely and the neural network classified the data as other
rhythm.

\section{\uppercase{Discussion}}
\label{sec:discussion}

\noindent
The neural network architecture we developed achieved a classification
performance similar to the winners of the challenge of Computing in Cardiology
2017. Although we could not evaluate our approach on the same data since the
true test set was not released yet, the results are promising. In particular,
the specificity for atrial fibrillation shows that only a limited number cases
are not detected. We could take into account signals of different lengths by
combining convolutional and recurrent layers. Indeed, the convolutional layers
extract features relevant for arrhythmia classification from sliding windows of
raw ECG signal. With this approach, there is no need to pre-process the data to
extract spectrogram or wavelet transform for instance. Furthermore, no feature
engineering is required as the neural network learns during the training phase
to extract high-level features useful for classification. The only
pre-processing we used is band-pass filtering as well as scaling to accelerate
training. We also used strategies for data augmentation to reduce over-fitting
and improve the generalization performance of the network. In particular,
randomly flipping the sign of ECG signals forced the neural network to learn to
take into account both regular and inverted waveforms. Without this simple
strategy, we would have needed to develop a robust method to detect signals
that had to be rectified.

In addition, we have shown that it is possible to apply a neural network
trained on a generic dataset to ECG signals recorded by the smart vest with
minimal changes. Indeed, there was no need to adapt the network architecture.
The method to standardize the ECG signals was the only element that required a
modification. Of course, if the two measurement systems, the AliveCor device
and the smart vest, did not record the same ECG lead, additional changes would
be needed. However, it is difficult to determine the extend of these changes
without evaluating the neural network on a dataset of ECG signals from another
lead.

Taken together, these results demonstrate that our system could be used to
monitor the cardiac activity of a subject for detecting abnormal rhythms over
long periods of time. Indeed, the smart vest is more comfortable to wear than
a traditional Holter and it can still collect high quality ECG signals since it
was initially developed for athletes. After transmitting the data to the cloud,
our neural network can quickly classify abnormal rhythms. While the accuracy
of our system is not perfect, it can still help to reduce the time spent
reviewing ECG by selecting segments with potential abnormalities that require
additional attention. These segments can then be analyzed by a trained
specialist. If needed, a full 12-lead ECG can be performed to refine or confirm
the diagnosis.

\section{\uppercase{Conclusion}}
\label{sec:conclusion}

\noindent
We presented a system composed of a smart vest to record a single lead ECG
signal and a neural network for detection and classification of heart
arrhythmias. We plan to aggregate several databases of ECG signals with rhythm
annotations to extend the types of arrhythmias that our algorithm can detect
and improve its accuracy. We will also investigate whether adding skip
connections in our network architecture further improves classification
performance.

\section*{\uppercase{Acknowledgements}}

\noindent
We would like to thank the anonymous reviewers for their valuable suggestions
and comments.

\bibliographystyle{apalike}
{\small
\bibliography{bibliography}}



\end{document}